\documentstyle[preprint,aps,eqsecnum]{revtex}

\newcommand{\lbar}{\overline}

\newcommand{\beq}{\begin{equation}}
\newcommand{\eeq}{\end{equation}}
\newcommand{\beqs}{\begin{eqnarray}}
\newcommand{\eeqs}{\end{eqnarray}}

\newcommand{\bhz}{\hat{\mbox{\boldmath $z$}}}

\newcommand{\bp}{{\mbox{\boldmath $p$}}}
\newcommand{\bk}{{\mbox{\boldmath $k$}}}
\newcommand{\bs}{{\mbox{\boldmath $s$}}}

\newcommand{\bsigma}{\mbox{\boldmath $\sigma$}}
\newcommand{\blambda}{\mbox{\boldmath $\lambda$}}

\newcommand{\sblambda}{\mbox{\boldmath \scriptsize$\lambda$}}
\newcommand{\Ep}{E_f}

\def\lsim{\ \rlap{\raise 3pt \hbox{$<$}}{\lower 3pt \hbox{$\sim$}}\ }
\def\gsim{\ \rlap{\raise 3pt \hbox{$>$}}{\lower 3pt \hbox{$\sim$}}\ }
\def\vev#1{\langle #1 \rangle}

\def\Tr{{\rm Tr}}

\def\sslash{s\!\!\!\slash}
\def\kslash{k\!\!\!\slash}

\def\pslash{p\!\!\!\slash}
\def\delslash{\partial\!\!\!\slash}

\def\half{{\textstyle{1 \over 2}}}
\def\ihalf{{\textstyle{i \over 2}}}
\def\eighth{{\textstyle{1 \over 8}}}

\def\npb#1{Nucl.\ Phys.\ {\bf B#1}}
\def\plb#1{Phys.\ Lett.\ {\bf B#1}}
\def\prd#1{Phys.\ Rev.\ {\bf D#1}}
\def\prl#1{Phys.\ Rev.\ Lett. {\bf#1}}

\def\zpc#1{Z.~Phys.\ {\bf C#1}}

\def\rmp{Rev.\ Mod.\ Phys.\ }
\def\reppp#1{Rep.\ Prog. Phys.\  {\bf #1}}

\def\mpla#1{Mod. Phys. Lett. A {\bf #1}}
\def\sjnp#1{Sov. J. Nucl. Phys. {\bf #1}}

\begin{document}
\draft

{\tighten
\preprint{\vbox{\hbox{WIS-99/11/Mar.-DPP}
                \hbox{SLAC-PUB-8083}
                \hbox{UdeA-PE-99/001}
                \hbox{hep-ph/9903517}
                \hbox{March, 1999}}}
\title{~\\~\\ Neutrino propagation in matter with general interactions}

\footnotetext{\scriptsize Research at SLAC is supported by the 
                          U.S. Department of Energy 
                          under contract DE-AC03-76SF00515}
\author{Sven Bergmann\,$^a$, Yuval Grossman\,$^b$ and Enrico Nardi\,$^c$}
\address{ \vbox{\vskip 0.truecm}
  $^a$Department of Particle Physics \\
  Weizmann Institute of Science, Rehovot 76100, Israel \\
\vbox{\vskip 0.truecm}
  $^b$Stanford Linear Accelerator Center \\
  Stanford University, Stanford, CA 94309, USA \\
\vbox{\vskip 0.truecm}
  $^c$Departamento de F\'\i sica \\
Universidad de Antioquia,  
A.A. {\it 1226}, \ Medell\'\i n,  \ Colombia }

\maketitle

\begin{abstract}%
We present a general analysis of the effective potential for neutrino
propagation in matter, assuming a generic set of Lorentz invariant
non-derivative interactions. We find that in addition to the known
vector and axial vector terms, in a polarized medium also tensor
interactions can play an important role.  We compute the effective
potential arising from a tensor interaction. We show that the
components of the tensor potential transverse to the direction of the
neutrino propagation can induce a neutrino spin-flip, similar to the
one induced by a transverse magnetic field.
\end{abstract} 
} % end tighten

\newpage

%%%%%%%%%%%%%%%%%%%%%%%%%%%%%%
%%%%%%%%%%%   I   %%%%%%%%%%%%
%%%%%%%%%%%%%%%%%%%%%%%%%%%%%%

\section{Introduction}
Neutrino physics currently provides the strongest experimental
evidence for physics beyond the Standard Model (SM).  The atmospheric
neutrino anomaly~\cite{AN} and the solar neutrino~\cite{SN} problem
are best explained by neutrino oscillations.

Neutrino oscillations occur when the produced neutrinos are not
eigenstates of the Hamiltonian that describes their propagation.  In
vacuum, this is the case if the flavor eigenstates are non-trivial
linear combinations of different mass eigenstates. This requires
massive neutrinos that mix. It is well known that the neutrino
propagation in matter can be very different from that in vacuum. The
crucial fact is that coherent interactions with the background give to
the neutrino an ``index of refraction" which depends on its flavor.
This is because normal matter, which contains only first generation
fermions, is flavor asymmetric. For example, for standard weak
interactions, only electron neutrinos can have charged current
interactions with the background electrons. Thus, in matter the
effective electron neutrino mass depends on the electron density and
is enhanced with respect to the other flavors. This allows for the
possibility of level crossing between different neutrino eigenstates
in matter. If the electron neutrinos are produced with an effective
mass above the level crossing (the ``resonance'') an adiabatic
transition through the resonance can induce a significant
amplification of neutrino oscillations. This is known as the
Mikheyev-Smirnov-Wolfenstein (MSW) effect~\cite{MSW}. If light sterile
neutrinos exist, then also neutral current interactions are important
since only the active neutrinos are subject to it~\cite{sterile}. In a
polarized medium the neutrino effective mass also depends on the
average polarization of the background, and on the angle between the
neutrino momentum and the polarization
vector~\cite{Nunokawa,spin-dep}.

Many extensions of the SM imply massive neutrinos. It is important to
stress that these new physics models often predict also new neutrino
interactions.  In this case the SM picture can be significantly
changed~\cite{NP-MSW,Bergmann,BergmannKagan}, since the neutrino
effective mass will depend on both the SM and the new interactions.
For example, non-universal interactions may give rise to matter
effects that distinguish between muon and tau neutrinos.  Lepton
flavor violating interactions can induce an effective mixing in
matter, allowing for a resonant conversion even in the absence of
vacuum mixing.  The two effects combined together could induce
neutrino flavor transitions even for massless neutrinos.

Most of the discussions of neutrino oscillations in matter are based
on the effective Hamiltonian
\beq \label{VAVA}
{\cal H}_{(V,A)} 
= {G_F \over \sqrt2} \sum_{a=V,A}  (\bar \nu\,\Gamma_a\,\nu)\, 
\left[\bar \psi_f\,\Gamma^a\, (g_a + g'_a
\gamma^5)\,\psi_f\right]\, + {\rm h.c.}\,, 
\eeq
where $\Gamma^V=\gamma^\mu$, $\Gamma^A=\gamma^\mu \gamma^5$, $\psi_f$
are the field operators for the background fermions $f=e, p, n, \nu$
and $g_a$, $g'_a$ are suitable coupling constants parametrizing the
strength of the interactions.

Clearly, the standard neutral current and the Fierz rearranged charged
current \linebreak $(V-A) \, (V-A)$ structures are included in
(\ref{VAVA}).  However, ${\cal H}_{(V,A)}$ describes in fact a larger
set of interactions.  For example, several models where neutrinos
couple to new heavy scalars (like supersymmetric models without
$R$-parity and left-right symmetric models) imply low energy effective
interactions of the form $\bar\nu (S \pm P) \psi_f \, \bar \psi_f (S
\mp P) \nu$ that, after Fierz rearrangment, are also accounted for by
(\ref{VAVA}).  The interactions in~(\ref{VAVA}) only induce
transitions between neutrinos of the same chirality.  Therefore the
couplings between different helicity states that would flip the
neutrino spin are suppressed by the ratio between the neutrino mass
and its energy, $m/E\,$, and can be safely neglected.  Thus the matter
effects induced by~(\ref{VAVA}) only allow for flavor transitions that
conserve the neutrino spin.  (Note that transitions into sterile
neutrinos~\cite{sterile} are no exception.  The sterile neutrino is a
SM singlet, but the state that is produced via oscillations has
negative helicity.)

In contrast, neutrino transitions induced by a magnetic field result
in a spin-flip~\cite{magneticvac}: a left-handed neutrino is rotated
into a right-handed one. The rate of this transition depends on the
neutrino magnetic moment and on the strength of the component of the
magnetic field orthogonal to the direction of the neutrino
propagation.  If the SM is extended just by introducing right-handed
neutrinos, the resulting neutrino magnetic moment is vanishingly
small, and spin-flipping transitions are negligible even for the
largest conceivable magnetic fields.  Therefore, spin-flipping
transitions can be relevant for solar or supernova neutrinos only in
the presence of new physics that induce a very large neutrino magnetic
moment.

While the couplings in~(\ref{VAVA}) account for the SM weak
interactions as well as for some new physics interactions they are
clearly not the most general ones. In this paper we systematically
study the effects of all Lorentz invariant non-derivative interactions
of neutrinos with the background fermions. Namely, we add
scalar~($S$), pseudoscalar~($P$) and tensor~($T$) interactions, to the
vector~($V$) and axial-vector~($A$) interactions in~(\ref{VAVA}). In
our analysis we reproduce the known results for $V$ and $A$
interactions~\cite{Nunokawa}. The $S$ and $P$ interactions that couple
states with opposite chirality but the same helicity are suppressed by
$m/E$ and therefore are negligible.  Our main result is that
transverse tensor interactions induce effects which are not helicity
suppressed, because they couple states of both opposite chirality and
opposite helicity. We find that in a polarized medium these
interactions can flip the neutrino spin coherently. The overall effect
depends on the strength of the interaction, on the density of the
background and on the average polarization of the medium. The physics
is similar to the electromagnetic spin-flip, however in this case
spin-flipping transitions can be effective even for a vanishing
neutrino magnetic moment. We note that an effective tensor potential
does not need to arise from a fundamental tensor interaction. It can
also result after Fierz reordering from some specific scalar and
pseudoscalar couplings of the neutrinos to the background fermions.

%%%%%%%%%%%%%%%%%%%%%%%%%%%%%%
%%%%%%%%%%   II   %%%%%%%%%%%%
%%%%%%%%%%%%%%%%%%%%%%%%%%%%%%

\section{Neutrino propagation in matter with general interactions}

In this section we derive the neutrino propagation equation in matter
in the presence of the most general pointlike and Lorentz invariant
four-fermion interaction with the background fermions ($f=e, p, n,
\nu$). That is, we generalize~(\ref{VAVA}) to
\beq \label{Hint}
{\cal H}_{\rm int} = 
{G_F \over\sqrt{2}} 
\sum_a (\bar \nu\,\Gamma^a\,\nu)\, 
\left[\bar \psi_f\,\Gamma_a\, (g_a + g'_a
\gamma^5)\,\psi_f\right]\, + {\rm h.c.}\,, 
\eeq
where
$\Gamma^a=\{I,\gamma^5,\gamma^\mu,\gamma^\mu\gamma^5,\sigma^{\mu\nu}\}$,
$\sigma^{\mu\nu}=\ihalf[\gamma^\mu,\gamma^\nu]$ and
$a=\{S,P,V,A,T\}$. Here the neutrino $\nu$ is assumed to be of the
Dirac type (we will comment on the Majorana case later).  In general,
$\nu$ is a vector of the different neutrino types, and $g_a, g'_a$ are
10 matrices in the space of neutrino flavors that describe the
coupling strengths.  In~(\ref{Hint}) the Fermi constant $G_F$ has been
factored out so that all the couplings are dimensionless.  From the
hermiticity of ${\cal H}_{\rm int}$ in~(\ref{Hint}) it follows that
all $g_a$ as well as $g'_V,g'_A$ are hermitian while $g'_S, g'_P$ and
$g'_T$ are antihermitian. In particular, the diagonal elements in
$g_a$ and $g'_V,g'_A$ are real while those of $g'_S, g'_P$ and $g'_T$
are imaginary.  We stress that new interactions in general include
both flavor diagonal and off-diagonal couplings.  The SM charged
current interactions of a $\nu_e$ with background electrons correspond
to $g_V=-g'_V=g_A=-g'_A = 1$ for the $\nu_e-\nu_e$ entries, while all
the other couplings vanish.

The derivation of the equation of motion describing the neutrino
propagation in a medium proceeds as follows. First we average the
effective interactions over the background fermions.  We are not
interested in incoherent effects that become negligible after
averaging.  Therefore, while we do allow for neutrino spin-flipping
interactions, we require that the background fermions do not undergo
spin-flip. That is, we select coherent transitions that leave the
many-fermion background system in the same state. Next we add the
effective neutrino interaction to the free Lagrangian, and we derive
the equation of motion for neutrino propagation in matter.  Finally,
we study the neutrino dynamics described by the equation of motion,
under the assumption that the masses and potential terms are much
smaller than the neutrino energy.

\subsection{Computing the effective neutrino potential} 
\label{ComputePotential}
The effect of the medium on the neutrino propagation in the presence
of the general interactions~(\ref{Hint}) can be described by the
Lagrangian
\beq \label{Lint} 
-{\cal L}_{\rm int} = \sum_{a,f} 
(\bar \nu\,\Gamma^a\,\nu)\, V^f_a \, , 
\eeq
where    
\beq \label{Va}
V_a^f = {G_F \over \sqrt{2}} 
\sum_{\sblambda} \int {d^3 p \over (2\pi)^3} 
\rho_f(\bp,\blambda)
{\cal M}_a^f \, ,
\eeq
is given by the expectation value of the background fermion current
${\cal M}_a^f$, averaged over the fermion distribution
$\rho_f(\bp,\blambda)$.  Here $\bp$ and $\blambda$ denote,
respectively, the momentum and polarization vectors of the background
fermion $f$. According to the requirement of leaving the many-fermion
background system unmodified, the matrix element
\beq \label{matrixelement}
{\cal M}_a^f \equiv 
\langle f,\bp,\blambda|\bar \psi_f\,\Gamma_a\,(g_a + g'_a
\gamma^5)\,\psi_f|f,\bp,\blambda 
\rangle 
\eeq
is taken between initial and final states with the same quantum
numbers. The computation of the various ${\cal M}_a^f$ is
straightforward and is given in the Appendix. We find
\beqs \label{Vgen}
V^S &=&      \label{VS}
{G_F \over \sqrt{2}} \, n_f \, g_S \left<{m_f \over E_f} \right> \, ,  \\
V^P &=&      \label{VP} 
{G_F \over \sqrt{2}} \, n_f \, g'_P \left< {m_f \over E_f} \right> \, ,  \\
V^V_\mu &=&  \label{VV}
{G_F \over \sqrt{2}} \, n_f 
\left[g_V \, \left<{p_\mu\over E_f}\right> 
    + g'_V \, m_f\, \left< {s_\mu \over E_f}\right> \right] \, , \\
V^A_\mu &=&  \label{VA}
{G_F \over \sqrt{2}} \, n_f 
\left[g'_A \, \left<{p_\mu\over E_f}\right> 
    + g_A \, m_f\, \left<{s_\mu \over E_f} \right> \right] \, , \\
V^T_{\mu\nu} &=&  \label{VT}  
{G_F \over \sqrt{2}} \, n_f
\left[-g_T \, \epsilon_{\mu\nu\rho\sigma} \, 
 \left<{p^\rho s^\sigma\over E_f} \right> 
+ i g'_T\,\left< {p_\mu s_\nu - p_\nu s_\mu\over E_f} \right> \right] \, , 
\eeqs
where the spin-vector $s\,$, which satisfies $s^2=-1\,$ 
and $s_\mu \, p^\mu=0\,$, is given explicitly in~(\ref{spinvector}), and
\beq
n_f = \sum_{\sblambda} \int {d^3 p \over (2\pi)^3} 
\rho_f(\bp,\blambda) \, , 
\qquad
\vev{x}={1 \over n_f}\sum_{\sblambda} \int {d^3 p \over (2\pi)^3} 
\rho_f(\bp,\blambda) \, x(\bp,\blambda) \,   
\eeq
denote, respectively, the number density of the fermion $f$ and the
average of some function $x(\bp,\blambda)$ over the fermion
distribution.

We can now perform the contractions $\Gamma^a \, V_a^f$
in~(\ref{Lint}), which yield
\beqs
\Sigma^{SP} &\equiv& \Sigma^0 \left[V^S + V^P \, \gamma^5 \right] ~= 
{G_F \over \sqrt{2}} \, n_f 
\left<{m_f \over E_f} \right> \, \left(g_S + g'_P \, \gamma^5 \right) 
\label{VSP}  \\
\Sigma^{VA} &\equiv& \gamma^\mu \left[V_\mu^V + V_\mu^A \, \gamma^5\right] ~= 
{G_F \over \sqrt{2}} \, n_f \left[
\left<{\pslash\over E_f}\right> \, \left(g_V  + g'_A \, \gamma^5 \right)
+ m_f \, \left< {\sslash \over E_f}\right> 
\left(g'_V + g_A \, \gamma^5 \right) \right] 
\label{VVA}  \\
\Sigma^T&\equiv& \Sigma^i\left[V^B_i + i V^E_i \, \gamma^5 \right]
= {G_F \over \sqrt{2}} \, n_f 
\left<{[\sslash, \pslash]\over E_f} \right> \, 
\left(g'_T + g_T \, \gamma^5 \right)
\label{VTT}  
\eeqs
where $\Sigma^\mu \equiv \mbox{diag} \, (\sigma^\mu, \sigma^\mu)$ with
$\sigma^\mu = (\sigma^0\,,\sigma^i)\,$ and $\sigma^0 = I\,$, and we
have used $\sigma^{ij}=\epsilon^{ijk} \, \Sigma^k$ and
$\sigma^{0i}=i\,\Sigma^i\gamma^5$. In~(\ref{VTT}) we have decomposed
the tensor term $V^T_{\mu\nu}$, in analogy to the electro-magnetic
field tensor $F_{\mu\nu}$, as $V^B_i = \epsilon_{ijk} \, V^T_{jk}$ and
$V^E_i = 2\,V^T_{0i} $.  Note that the second equality in~(\ref{VTT})
makes apparent that the tensor interaction can contribute only in the
presence of a polarized background.

\subsection{Equations of Motion}
We turn now to study the effects of the potential on the neutrino
propagation.  The equation of motion can be deduced from the neutrino
Lagrangian
\beq \label{Lagr}
{\cal L} = {\cal L}_{free} + {\cal L}_{int}=
\bar{\nu} (i \delslash - m - \Sigma) \nu 
\eeq
where  the matrix of the potentials 
\beq
\Sigma  \equiv \Sigma^{SP} + \Sigma^{VA} +  \Sigma^T\, 
\eeq
depends on the background density and polarization, and in general
will vary along the neutrino propagation path.  In the general case
both $\Sigma$ and $m$ are matrices in the space of neutrino types.
It is instructive to write the interaction part in~(\ref{Lagr})
explicitly in the chiral basis, see~(\ref{chiralbasis})
\beq \label{Lint2}
-{\cal L}_{int} = \bar \nu \, \Sigma \, \nu =
\pmatrix{\nu^\dagger_L \cr \nu^\dagger_R}^T
\pmatrix{V^{LL}_\mu \bar\sigma^\mu &  V^{LR}_\mu \sigma^\mu \cr
         V^{RL}_\mu     \sigma^\mu &  V^{RR}_\mu \sigma^\mu \cr}
\pmatrix{\nu_L \cr \nu_R},
\eeq
where $\bar\sigma^\mu = (\sigma^0\,, - \sigma^i)$ and 
\beqs
V^{LL}_\mu &\equiv& V_\mu^V - V_\mu^A\,,~~~~~~~~~
V^{RR}_\mu \equiv V_\mu^V + V_\mu^A\,, \label{LLVA} \\
V^{RL}_0 &\equiv& V^S - V^P\,,~~~~~~~~~ 
V^{LR}_0 \equiv V^S + V^P\,, \label{LRSP} \\
V^{RL}_i &\equiv& V_i^B-i\,V_i^E\,,~~~~~~~   
V^{LR}_i \equiv V_i^B+i\,V_i^E\,.  \label{LRT} 
\eeqs
The explicit form~(\ref{Lint2}) makes apparent that the (axial)vector
potentials (contained in $V^{LL}$ and $V^{RR}$) couple neutrinos of
the same chirality, while the (pseudo)scalar and tensor potentials (in
$V^{RL}$ and $V^{LR}$) couple neutrinos of opposite chirality.

From~(\ref{Lagr}) it follows that the equations of motion for
neutrinos and antineutrinos are, respectively,
\beq \label{EOMnu}
\gamma_0 (\kslash - m - \Sigma) u = 0\,, \qquad\quad 
\gamma_0 (\kslash + m + \Sigma) v = 0\,.   
\eeq
We note that the signs of $m$ and $\Sigma$ are opposite for the
antineutrinos. The dispersion relations for the neutrino propagation
are given by the solutions of
\beq \label{Disp} 
{\rm det}\,[{\cal O}] = {\rm det}\,[\gamma_0(\kslash - m -\Sigma)] = 0. 
\eeq 
Solving~(\ref{Disp}) is simplified by working in the following
approximation. Let us chose the neutrino momentum along the $z$-axis
($\bk = k \bhz\,$). Then $\sigma_{0,3}$ couple between states of the
same helicity while $\sigma_{1,2}$ couple neutrinos of opposite
helicity. Hence, for ultra-relativistic neutrinos, $V^{LL}_{1,2}$ and
$V^{RR}_{1,2}$ in the chirality conserving diagonal blocks
in~(\ref{Lint2}) and $V^{LR}_{0,3}$ and $V^{RL}_{0,3}$ in the
chirality flipping off-diagonal blocks are suppressed as $m/E \ll 1$,
and can be neglected. Thus, the relevant potential terms
in~(\ref{Lint2}) are $V^{LL}_{0,3}$, $V^{RR}_{0,3}$ and the tensor
potential components $V^{LR}_{1,2}$ and $V^{RL}_{1,2}$ that are
transverse with respect to the neutrino propagation direction. In this
approximation we get
\beq \label{fullmatrix}
{\cal O} =
\pmatrix{E+ k - V^{LL}_{0+3} & 0 & -m & -V^{LR}_- \cr
         0 & E - k - V^{LL}_{0-3} & -V^{LR}_+ & -m \cr
         -m & -V^{RL}_- & E - k - V^{RR}_{0-3} & 0 \cr
         -V^{RL}_+ & -m & 0 & E + k - V^{RR}_{0+3} } \, ,
\eeq
where $V_{0 \pm 3} \equiv V_0 \pm V_3\,$ and $V_\pm \equiv V_1 \pm i
V_2$. Note that since $V^{V,A,T}$ are hermitian, $(V^{RL}_\pm)^\dagger
= V^{LR}_\mp\,$ and the matrix~(\ref{fullmatrix}) is manifestly
hermitian. Solving the determinant equation for~(\ref{fullmatrix})
under the assumption that $V^{V,A,T}, m \ll E\,$ yields the neutrino
energies:
\beq \label{Solution}
E_\pm  =
k + {m^2\over 2 k}+ \half \left[V^{LL}_{0-3} + V^{RR}_{0-3} \pm 
\sqrt{\left(V^{LL}_{0-3} - V^{RR}_{0-3}\right)^2 + 
      4\, V^{LR}_+ \, V^{RL}_-} \, \right] \,,   
\eeq
where the plus (minus) sign refers to neutrinos that are mainly
left(right)-handed states.  Eliminating the two helicity suppressed
states from the equations of motion we obtain a Schr\"odinger-like
equation that governs the neutrino propagation:
\beq \label{EOMLR}
i {d \over dt} \pmatrix{\nu_L \cr \nu_R} = 
{\cal H}_\nu \pmatrix{\nu_L \cr \nu_R} ~~~~
\mbox{with}~~~~{\cal H}_\nu =
k + {m^2 \over 2k} +
\pmatrix{V^{LL}_{0-3} & V^{LR}_+ \cr 
         V^{RL}_-     & V^{RR}_{0-3} \cr} \,.
\eeq
The two energy eigenvalues of the effective Hamiltonian ${\cal H}_\nu$ are
the solutions~(\ref{Solution}) of the determinant equation~(\ref{Disp}).
The equation for the antineutrinos, and the corresponding eigenvalues,
can be obtained from~(\ref{EOMLR}) and~(\ref{Solution}) by changing the
sign of the potentials ($V \to -V$). Note that the contribution to the
energy levels from the tensor term, which is quadratic in $V^T$, does
not change sign. In the case of more than one neutrino
flavor~(\ref{Solution}) is a matrix equation in the space of the
neutrino types. It is interesting to note that in general we should
not expect that the various interactions in~(\ref{Solution}) will be
diagonal in the same basis. In this case even in the massless limit
(or for degenerate neutrinos) flavor oscillations can occur in
matter. In the one flavor case, the energy gap between the two states
is
\beq \label{DeltaE}
\Delta E_\nu =
\sqrt{\left(V^{LL}_{0 - 3} - 
V^{RR}_{0 - 3} \right)^2 + 4\, V^{LR}_+ \, V^{RL}_-} \, . 
\eeq
In the limit of vanishing tensor interaction $(V_T=0)\,$ $\nu_L$
decouples from $\nu_R\,$, and we obtain
\beq \label{Enu1}
E_L = k + {m^2 \over 2k} +  V^{LL}_{0-3}, \quad   
\qquad
E_R = k + {m^2 \over 2k} +  V^{RR}_{0-3}. 
\eeq
Clearly in this case we can have oscillations only between different
neutrino flavors.  Moreover, if there is a basis where the full
$V^{LL}$ (or $V^{RR}\,$, for the SM sterile states) is flavor
diagonal, then oscillations can occur only in the presence of
non-trivial mixings in the mass matrix.  Setting $V^{RR}=0$ and
$V^{LL}$ equal to the SM charged current and neutral current
interactions, we recover the SM case, with non-interacting
right-handed states.

So far we only discussed the case of neutrinos propagating in a
background of particles. If also antiparticles (e.g. positrons) are
present in the background, one has to take into account the
corresponding interactions. Assuming CP conservation we find that
neutrino scattering off antifermions leads to the Hamiltonian
in~(\ref{EOMLR}), but with opposite sign for the potential matrix.

%%%%%%%%%%%%%%%%%%%%%%%%%%%%%%
%%%%%%%%%%   III   %%%%%%%%%%%
%%%%%%%%%%%%%%%%%%%%%%%%%%%%%%

\section{Implications and Discussion}
\label{discussion}

The general interactions that we have studied in the previous section
can give rise to several effects for neutrino oscillations in
matter. It is well-known that the vector and axial-vector interactions
can be very important for neutrino propagation in dense matter. These
interactions do not change the neutrino spin, but they can enhance
flavor transitions when the neutrino moves through a
resonance~\cite{MSW}.

To recover the SM result for the potential felt by an electron
neutrino propagating in an electron background, we set
$g_V=-g'_A=g_A=-g'_V=1\,$ in~(\ref{VVA}) and obtain
\beq \label{LintSM}
\Sigma^{SM} = \sqrt{2} \, G_F \, n_e
\left(\left<\pslash \over E_e\right>  -  m_e\,
\left<{\sslash \over E_e}\right>
\right) \, P_L \label{VSM} \,,
\eeq
with $P_L=\half(1-\gamma_5)\,$. Defining $\hat \bk=\bk/|\bk|$ as a
unit vector in the direction of the neutrino momentum $\bk$ and using
the explicit expression~(\ref{spinvector}) for the spin vector $s$ we
obtain
\beq \label{SMresult} 
V_{\nu,\bar\nu}^{SM} = \pm \sqrt{2} \, G_F \, n_e 
\left[1 
- \left<{\hat {\bk} \cdot \bp\over E_e}\right>  
- \left<{\bp\cdot\blambda \over E_e} \right> 
+m_e \left<{\hat\bk \cdot\blambda \over E_e} \right> 
+\left<{(\hat \bk\cdot\bp) \, (\bp\cdot\blambda) 
\over E_e(m_e+E_e)}\right> \right] \, ,
\eeq
which is valid for an arbitrary neutrino direction.  The plus-sign
in~(\ref{SMresult}) refers to neutrinos and the minus-sign to
antineutrinos. We note that~(\ref{SMresult}) is in agreement with the
results given in~\cite{Nunokawa}.

Our main result is, however, that in the presence of a neutrino tensor
interaction with the background fermions, the neutrino can undergo
spin-flip. This effect is similar to the spin-precession induced by a
transverse magnetic field $B_\perp$ that couples to the neutrino
magnetic dipole moment $\mu_\nu\,$. In fact, if we substitute
in~(\ref{EOMLR}) the off-diagonal term $V^{LR}_\pm$ by $\mu_\nu
B_\perp$ we obtain the equation of motion for a neutrino that
propagates in a magnetic field~\cite{magneticvac,magneticmat}. Thus,
while these two scenarios originate from different physics, formally
they can be treated in the same way.

To illustrate the effects of neutrino oscillations due to the presence
of a non-zero transverse tensor potential, we consider the simplest
case of one neutrino generation. A left-handed neutrino that was
produced at $t=0$ and propagates for a time $t$ in a constant medium
will be converted into a right-handed neutrino with a probability
\beq
P_{\nu}^{LR}(t) = 
\sin^2 2\theta \, 
\sin^2 \left({\Delta E_{\nu} \, t \over 2} \right) \,.
\eeq
The effective mixing angle $\theta$ is given by
\beq \label{mixing}
\sin^2 2\theta = {|2 V^{LR}_+|^2 \over (\Delta E_{\nu})^2}\,,
\eeq
where the energy splitting $\Delta E_{\nu}$ is defined
in~(\ref{DeltaE}). (Note that for one neutrino flavor we have
$V^{LR}_+ \,V^{RL}_- = |V^{LR}_+|^2\,$.) In the case of more than one
neutrino flavor, propagation in a medium with changing density can
lead to resonance effects in complete analogy to the magnetic field
induced resonant spin-flip. We will not discuss the details of the
resonant case here (which can be found in the existing
literature~\cite{magneticmat}), but we want to discuss shortly the
results for different types of background matter.

First consider a medium where the average momentum of the background
fermions vanishes ($\vev{\bp} = 0$). This is in particular the case
for an isotropic momentum distribution. Then, the relevant
(transverse) component of the tensor potential which determines the
effective mixing in~(\ref{mixing}) is given by
\beq \label{isotropic} 
|V^{LR}_+| = \sqrt{2} \, G_F \, n_f
\sqrt{|g_T|^2 + |g_T'|^2} \, \left< \lambda_\perp\> \left(
\sin^2\vartheta+ {m_f \over E_f}
\cos^2\vartheta\right)
\right>\,,
\eeq
where $\vartheta$ is the angle between the momentum and the transverse
polarization of the of the background fermion and
$\lambda_\perp=\sqrt{\lambda_1^2 + \lambda_2^2}\,$.  Note that
$|V^{LR}_+|$ vanishes if the neutrino propagates along the direction
of the average background polarization ($\lambda_\perp=0$). For a
non-relativistic background (where $E_f \simeq m_f \gg p_i$) we obtain
from~(\ref{isotropic}) that the effective mixing angle is determined
by
\beq \label{nonrel}
|V^{LR}_+| = \sqrt{2}\,G_F \, n_f
\sqrt{|g_T|^2 + |g_T'|^2} \, \vev{\lambda_\perp} \,.
\eeq
In the ultra-relativistic limit the effective mixing depends on
$\vev{\lambda_\perp\,\sin^2\vartheta}$ which is equal to
$\vev{\lambda_\perp/2}$ if $\lambda_\perp$ is uncorrelated to the
momentum of the background fermion. Finally, for a degenerate
background in the presence of a magnetic field, only the fermions in
the lowest Landau level contribute to the polarization, with the spin
oriented antiparallel to the momentum. In this case the background is
not isotropic, and eq.~(\ref{isotropic}) is not applicable.  One
obtains for this case
\beq \label{landau}
|V^{LR}_+| = \sqrt{2}\,G_F \, n_f
\sqrt{|g_T|^2 + |g_T'|^2} \, 
\left<\lambda_\perp {m_f \over E_f}\right> \,,
\eeq
which vanishes in the ultra-relativistic limit.

Let us now comment on the possible source of the tensor interaction.
Of course, one cannot rule out elementary tensor interactions.
However, it is interesting to note that also certain neutrino scalar
interactions can generate, after Fierz rearrangement, effective tensor
couplings. For example, consider the tree level Lagrangian
\beq \label{phiint}
-{\cal L}_{\rm tree}=
\lambda_\phi \phi \, (\lbar{L_L} \, e_R) +
\lambda'_\phi \tilde\phi \, (\lbar{L_L} \, \nu_R) \, + {\rm h.c.}\,, 
\eeq 
where $L_L$ is the left-handed lepton $SU(2)_L$ doublet, $e_R$
($\nu_R$) is the right-handed electron (neutrino) singlet, $\phi$ is a
doublet scalar field, of mass $m_\phi$ and $\tilde\phi = i \sigma_2
\phi^*$ and $\lambda_\phi, \lambda'_\phi$ are real elementary
couplings. At low energy $E \ll m_\phi$, the interaction
in~(\ref{phiint}) induces a set of four-fermion effective
interactions, which also contains the following coupling
\beq \label{fourfermi}
{\cal H}_{\rm int}^\phi = {\lambda'_\phi \lambda_\phi \over m_\phi^2} 
(\lbar{e_R} \, \nu_L) \, (\lbar{\nu_R} \, e_L) = 
-{\lambda'_\phi \lambda_\phi \over m_\phi^2} 
\left[
\half(\lbar{\nu_R} \, \nu_L) \, (\lbar{e_R} \, e_L) + 
\eighth
(\lbar{\nu_R} \, \sigma_{\mu \nu} \, \nu_L) \, 
(\lbar{e_R} \, \sigma^{\mu \nu} \, e_L)
\right] \, .
\eeq 
From~(\ref{fourfermi}) it follows that $g_T \sim {\lambda'_\phi
\lambda_\phi/ m_\phi^2}$.  Finally, we mention that the above
four-fermion operator can also be generated when different scalar
fields mix.  This possibility exist, for example, in supersymmetric
models without $R$-parity.

Throughout this paper we assumed the neutrinos to be of the Dirac
type. For the case of Majorana neutrinos there are additional
constraints on some of the couplings. Namely, one can
show~\cite{majorana} that the flavor diagonal elements of the vector
couplings $g_V, g'_V\,$ as well as the tensor couplings $g_T, g'_T\,$
vanish identically, while the axial-vector couplings are twice the
value corresponding to the Dirac case. As a consequence the standard
MSW effect does not distinguish between Dirac and Majorana neutrinos,
but a tensor-induced spin-flip requires at least two neutrino flavors
in the Majorana case.

Let us now address shortly the issue whether the new tensor term could
be relevant for real physical systems, like the Sun or a supernova.
The crucial point is that the effective tensor potential is
proportional to the tensor couplings and to the average background
polarization. From eqs.~(\ref{isotropic})--(\ref{landau}) it follows
that in general it is suppressed by a factor
\beq \label{eps}
\epsilon \equiv \left| V^{LR}_+ \over V^{LL}_0 \right|
\lsim \sqrt{|g_T|^2 + |g_T'|^2} \, \langle \lambda_\perp \rangle
\eeq
with respect to the SM vector potential.  New physics effects can be
relevant to neutrino oscillations only if they are large enough to
affect sizably the standard results obtained with the usual SM
interactions.  The problem of estimating the minimum size required to
render these effects observable was addressed
in~\cite{Bergmann,BergmannKagan}.  These analyses imply that the
tensor interaction could be relevant respectively for solar and
supernova neutrino oscillations, if $\epsilon$ satisfies the following
lower limits:
\beq
\epsilon_{sun} \gsim 10^{-2} \,~~~~\mbox{and}~~~~
\epsilon_{SN} \gsim 10^{-4} \,. 
\eeq
According to~(\ref{eps}) the effect is maximal for the maximum allowed
values of $g_T, g_T'$ and $\langle \lambda_\perp \rangle$. Clearly,
the excellent agreement between the SM predictions for processes
involving neutrinos and the corresponding experimental results,
suggests that the tensor couplings are small.  We expect that,
besides the direct limits from decays and from neutrino scattering
data, in some cases one can also derive severe constraints from
$SU(2)_L$ related interactions~\cite{Bergmann-Grossman} as well as
from the bounds on neutrino masses.  While a detailed phenomenological
analysis is needed to give definite upper bounds on the tensor
couplings~\cite{WorkInProgress}, we believe that they will not exceed
the few percent level.

However, the suppression of the tensor potential due to the average
polarization is by far the most important factor.  In the solar
interior, the magnetic field can be at most of the order of several
kG. This can result in a tiny polarization of the (non-relativistic)
electrons,
\beq
\vev{\lambda_{e}} \sim {eB\over m_e T}   \simeq 10^{-8} \left[B\over 1\, 
{\rm kG}\right]
\left[1\, {\rm keV}\over T\right]\,,  
\eeq
where $B$ and $T$ denote, respectively, the magnetic field and the
temperature in the relevant region of propagation.  We conclude that
quite likely neutrino propagation in the Sun cannot be affected by the
new tensor interactions.

In a proto-neutron star during the early cooling phase, a few seconds
after the supernova explosion, the magnetic field strength can reach
extremely large values.  However, the temperature is also large, thus
suppressing the induced polarization.  Since the electron number
density is only about 10\% of the nucleon density, and due to the fact
that the electrons are relativistic and degenerate, the effect of
neutron polarization can be comparable, and even dominant, with
respect to the effect of electron polarization. We estimate
\beq\label{polarization}
\vev{\lambda_{p,n}} \simeq 10^{-5} 
\left[B\over 10^{13}\, {\rm G}\right]
\left[10\, {\rm MeV}\over T\right] 
~~~\mbox{and}~~~~
\vev{\lambda_{e}} \simeq 10^{-4} 
\left[B\over 10^{13}\, {\rm G}\right]
\left[20\, {\rm MeV}\over k_F\right]^2 \, , 
\eeq
where $k_F$ is the Fermi momentum of the degenerate electrons.  The
above suggests that for conservative values of the magnetic field, of
the order of $B\lsim 10^{13}$\,G, it is unlikely that the tensor
interaction could affect the propagation of supernova neutrinos.
However, one cannot rule out completely the possibility of large
enhancements of the effects of the tensor interaction. First, inside
the supernova core the neutrinos are not freely streaming and suffer
collisions. In general, the effect of collisions is to increase the
production of the right-handed states. Second, the magnetic field
inside the core is poorly known.  It has been proposed that at early
times inside the proto-neutron star the magnetic field could be as
strong as $10^{16}\, $G~\cite{strongB}.  This would imply an
enhancement of the polarization of about three orders of magnitude,
opening the possibility of observing these effects. Finally, it has
also been speculated that very dense and neutron rich matter could
have a ferromagnetic phase~\cite{ferromagnetic} (even if this is
unlikely to occur at the time of neutrino emission, when the
temperature is very high, there could still be some large enhancement
of $\vev{\lambda_{p,n}}$).  Also in this case neutrino tensor
interactions with the highly polarized background could be effective
for inducing transitions into right-handed states at a sizable rate.
Of course, since the presence of right-handed neutrinos implies in
general a non-vanishing magnetic moment, the effect of the tensor
interaction will be accompanied by a similar effect of the neutrino
magnetic moment coupled to the strong magnetic field. In this case,
both effects have to be taken into account simultaneously. This and
related issues will be discussed elsewhere~\cite{WorkInProgress}.

To conclude, in this paper we have studied the effects on neutrino
propagation in matter due to the most general Lorentz-invariant
interactions with the background fermions.  Scalar, pseudo-scalar and
longitudinal tensor interactions couple states of opposite chirality
but do not flip the helicity, and hence are suppressed by the ratio
between the neutrino mass and its energy. Our crucial observation is
that transverse tensor interactions are not suppressed by this ratio,
since they couple states of both opposite chirality and opposite
helicity and they can be coherently enhanced in the presence of a
non-vanishing background polarization. As a result, such interactions
can induce a neutrino spin-flip during propagation, much alike the
magnetic moment spin-precession~\cite{magneticvac,magneticmat}.

Applying our scenario to astrophysical neutrino sources, we find that
the suppression from the average background polarization and the
tensor couplings, implies that this effect is probably irrelevant for
solar neutrinos. For supernova neutrinos the effect could become
observable only in the presence of extremely large magnetic fields
or, more speculatively if some new mechanism can enhance by a few orders
of magnitude the conversion rate. Clearly, a definite conclusion about
the relevance of this effect for different physical systems requires
further investigation~\cite{WorkInProgress}.

\acknowledgements 
We thank L. Dixon, H. Lipkin, Y. Nir, S. Nussinov, C. Quigg, A. Stern
and N. Weiss for helpful discussions.  Y.G. is supported by the
U.S. Department of Energy under contract DE-AC03-76SF00515.

%%%%%%%%%%%%%%%%%%%%%
%%%   Appendix    %%%
%%%%%%%%%%%%%%%%%%%%%

\newpage
\appendix
\section{}
We present here the details of the computation of the matrix elements 
${\cal M}_a$ [c.f.~(\ref{matrixelement})] that determine the potentials
$V_a$ [c.f.~(\ref{Va})]. We have 
\beqs 
{\cal M}_a^f &\equiv& 
\langle f,\bp,\blambda|\lbar \psi_f\,\Gamma_a\,(g_a + g'_a
\gamma^5)\,\psi_f|f,\bp,\blambda \rangle \\
&=&   \label{matrixspinor}
{\textstyle{1 \over 2 E_f}}
\lbar u_f(\bp,\blambda) \, \Gamma^a\,(g_a + g'_a \gamma^5) \,
u_f(\bp,\blambda) \\
&=&   \label{matrixtrace}
{\textstyle{1 \over 4 E_f}}
{\rm Tr}\left[\Gamma^a\,(g_a + {g_a}' \gamma^5) \,
(\pslash + m_f) \, (1+\gamma^5\sslash)\, \right] \, ,
\eeqs 
where $E_f$ and $m_f$ denote respectively the energy and the mass of
the background fermion $f$. In~(\ref{matrixspinor}) we have assumed
the background fermions to be free, so that a plane wave expansion for
the field operators can be used. In obtaining~(\ref{matrixtrace}) we
have used the identity
\beq \label{trace}
u_f(\bp,\blambda) \, \lbar u_f(\bp,\blambda) = \half (\pslash +
m_f) \, (1+\gamma^5\sslash) \, ,
\eeq
where the spin vector $s$ is defined as
\beq \label{spinvector}
s \equiv \left({\bp\cdot\blambda \over m_f}, \blambda + 
{\bp \,  (\bp\cdot\blambda) \over m_f\, (m_f+E_f)}\right) \, ,
\eeq
and satisfies $s^2=-1$ and $s_\mu \,p^\mu=0$. Using $\gamma^5
\sigma^{\mu\nu} = {i\over 2} {\epsilon^{\mu\nu}}_{\rho\sigma} \,
\sigma^{\rho\sigma} $ and the elementary traces ${1\over 4} \, \Tr\,
[\Gamma^{S,P,V,A,T} (\pslash + m_f)(1+\gamma^5 \sslash)] = m_f \,, 0
\,, p^\mu \,, s^\mu \,, -\epsilon^{\mu\nu\rho\sigma} \, p_\rho
s_\sigma$ we obtain
\beqs 
{\cal M}^S &=&  
g_S \,  { m_f \over E_f} \, ,
\label{MS} \\
{\cal M}^P &=&  
g'_P  \, {m_f\over E_f} \, ,
\label{MP} \\
{\cal M}^V &=&  
g_V \, {p^\mu\over E_f} +  g'_V \, {m_f\over E_f}\, s^\mu \, ,
 \label{MV} \\
{\cal M}^A &=&  
g'_A \,{p^\mu \over E_f}+ g_A  \, {m_f\over E_f} \, s^\mu \, ,
\label{MA} \\
{\cal M}^T &=& - 
g_T \, \epsilon^{\mu\nu\rho\sigma} \, {p_\rho 
s_\sigma  \over E_f}
+  i g'_T \, {p^\mu s^\nu - p^\nu  
s^\mu\over E_f} \, . \label{MT}
\eeqs
While the identity~(\ref{trace}) provides a simple way to calculate
${\cal M}_a$ by means of standard ``trace technology'', we find it
useful to present also an alternative calculation which is based on
the spinorial expression~(\ref{matrixspinor}) for ${\cal M}_a$. In
this derivation the details of the fermion polarization
\beq
\blambda = \xi_f^\dagger \, \bsigma \, \xi_f ~~~
(\xi_f^\dagger \, \xi_f = 1) 
\eeq
are more transparent ($\xi_f$ denotes the two-component spinor of the
fermion $f$).

To compute $\lbar u(\bp,\blambda) \, \Gamma_a \, (g_a + g'_a \gamma^5)
\, u(\bp,\blambda)$ we choose the chiral representation for
$\Gamma_a$, where
\beq \label{chiralbasis}
1          = \pmatrix{I & 0 \cr 0 & I}, ~~~~~~
\gamma^5   = \pmatrix{-I & 0 \cr 0 & I}, ~~~~~~
\gamma^\mu = \pmatrix{0 & \sigma^\mu \cr \bar \sigma^\mu & 0},
\eeq
with $\sigma^\mu=(I,\sigma^i)$ and $\bar\sigma^\mu=(I,-\sigma^i)\,$.  
Since 
\beq \label{chiralspinor}
u(\bp,\blambda) \equiv \pmatrix{u_L(\bp,\blambda) \cr u_R(\bp,\blambda)} =
\pmatrix{\sqrt{p \, \sigma} \, \xi_f \cr \sqrt{p \, \bar\sigma} \, \xi_f}
= \sqrt{\Ep+m_f \over 2} 
\pmatrix{\left(I - {\bp\cdot\bsigma \over {\Ep + m_f}}\right) \, \xi_f \cr
         \left(I + {\bp\cdot\bsigma \over {\Ep + m_f}}\right) \, \xi_f } \,,
\eeq
it is sufficient to calculate $u_{C}^\dagger(\bp,\blambda) \,
\sigma^\mu \, u_{C'}(\bp,\blambda)$ for $C, C' \in \{L,R\}$. Using the
identities
\beqs
(\bp\cdot\bsigma) \sigma_i + \sigma_i (\bp\cdot\bsigma) &=& 2 p_i \\
(\bp\cdot\bsigma) \sigma_i - \sigma_i (\bp\cdot\bsigma) &=& 
 2i \epsilon_{kji} \sigma_k p_j \\
(\bp\cdot\bsigma) \sigma_i (\bp\cdot\bsigma) &=& 2 p_i (\bp\cdot\bsigma) -
 |\bp|^2 \sigma_i
\eeqs
we obtain
\beqs
u_{L,R}^\dagger(\bp,\blambda) \, I \, u_{L,R}(\bp,\blambda) &=& 
 (\Ep \mp \bp\cdot\bsigma) \\
u_{L,R}^\dagger(\bp,\blambda) \, I \, u_{R,L}(\bp,\blambda) &=& 
 m_f \\
u_{L,R}^\dagger(\bp,\blambda) \, \bsigma \, u_{L,R}(\bp,\blambda) &=& 
 m_f \blambda + \left({\bp (\bp\cdot\bsigma) \over {\Ep+m_f}} \mp \bp \right)
\\
u_{L,R}^\dagger(\bp,\blambda) \, \bsigma \, u_{R,L}(\bp,\blambda) &=& 
\Ep \blambda \pm i (\bp \times \blambda) 
- {\bp (\bp\cdot\bsigma) \over {\Ep+m_f}} \, .
\eeqs
This allows us to compute 
\beq
J_a \equiv \lbar u(\bp,\blambda) \, \Gamma_a \, u(\bp,\blambda)
\eeq
for $a = S,P,V,A$:
\beqs
J_S &=& u_L^\dagger u_R + u_R^\dagger u_L = 2 m_f \\ 
J_P &=& u_L^\dagger u_R - u_R^\dagger u_L = 0 \\
J_V &=& u_R^\dagger \sigma_\mu u_R +
        u_L^\dagger \bar\sigma_\mu u_L    = 2 p_\mu \\
J_A &=& u_R^\dagger \sigma_\mu u_R -
        u_L^\dagger \bar\sigma_\mu u_L    = 2 m_f s_\mu \, ,
\eeqs
where $s_\mu$ is defined in~(\ref{spinvector}).  From the above one
immediately obtains ${\cal M}_a$ for $a = S,P,V,A$ as
in~(\ref{MS}-\ref{MA}). To compute the tensor terms we define
\beq
\Sigma_i  \equiv \pmatrix{\sigma_i & 0 \cr 0 & \sigma_i}, ~~~~~~
\Sigma_i' \equiv \Sigma_i \, \gamma^5 =
 \pmatrix{-\sigma_i & 0 \cr 0 & \sigma_i}. 
\eeq
The respective currents are
\beqs
J_{\Sigma}  &=& u_L^\dagger \bsigma u_R +
                u_R^\dagger \bsigma u_L    
  = 2 \Ep \blambda - {{2 \bp (\bp\cdot\blambda)} \over {\Ep+m_f}} 
  = 2 \Ep \bs - 2 \bp s_0 \\ 
J_{\Sigma'} &=& u_L^\dagger \bsigma u_R -
                u_R^\dagger \bsigma u_L   
  = 2i (\bp \times \blambda)
  = 2i (\bp \times \bs)\,.
\eeqs
Noting that 
$
\sigma_{ij}          = \epsilon_{ijk} \Sigma_k \, ,~
\sigma_{ij} \gamma_5 = \epsilon_{ijk} \Sigma_k' \, ,~ 
\sigma_{0i}          = i \Sigma_i'~$and$~
\sigma_{0i} \gamma_5 = i \Sigma_i 
$
one can easily verify the expression for ${\cal M}_T$ in eq.~(\ref{MT}).

%%%%%%%%%%%%%%%%%%%%%%
%%%   References   %%%
%%%%%%%%%%%%%%%%%%%%%%

\end{document}